\documentclass[12pt,preprint]{aastex}
\usepackage[]{natbib}
\usepackage[]{graphicx}

\newcommand\lsun{\rm L_{\odot}}

\newcommand\be{\begin{equation}}
\newcommand\en{\end{equation}}

\shorttitle{HBC722 Submm Environs}
\shortauthors{Green et al.}

\begin{document}

\title{Disentangling the Environment of the FU Orionis Candidate HBC 722 with Herschel\thanks{{\it Herschel} is an ESA space observatory with science instruments provided by European-led Principal Investigator consortia and with important participation from NASA.}}

\author{Joel D. Green\altaffilmark{1}, Neal J. Evans II\altaffilmark{1}, \'{A}gnes K\'{o}sp\'{a}l\altaffilmark{2}, Tim A. van Kempen\altaffilmark{3,2},  Gregory Herczeg\altaffilmark{4}, Sascha  P. Quanz\altaffilmark{5},  Thomas Henning\altaffilmark{6}, Jeong-Eun Lee\altaffilmark{7}, Michael M. Dunham\altaffilmark{8}, Gwendolyn Meeus\altaffilmark{9}, Jeroen Bouwman\altaffilmark{6},  Ewine van Dishoeck\altaffilmark{2,4},  Jo-hsin Chen\altaffilmark{10}, Manuel G\"udel\altaffilmark{11}, Stephen L. Skinner\altaffilmark{12}, Manuel Merello\altaffilmark{1}, David Pooley\altaffilmark{1,13}, Luisa M. Rebull\altaffilmark{14}, \& Sylvain Guieu\altaffilmark{15}}

\affil{1. Department of Astronomy, University of Texas at Austin, TX \\
2. Leiden Observatory, Leiden University, Netherlands \\
3. Joint ALMA offices, Av. Alsonso de Cordova, Santiago, Chile \\
4. Max Planck Institute for Extraterrestrial Physics, Garching, Germany \\
5. Institute for Astronomy, ETH, Zurich, Switzerland \\
6. Max Planck Institute for Astronomy, Heidelberg, Germany \\
7. Department of Astronomy \& Space Science, Kyung Hee University, Gyeonggi 446-701, Korea  \\
8. Dept. of Astronomy, Yale University, New Haven, CT \\
9. Universidad Autonoma de Madrid, Spain \\
10. Jet Propulsion Laboratory, Pasadena, CA \\
11. Dept. of Astronomy, University of Vienna, Austria \\
12. Center for Astrophysics and Space Astronomy (CASA), University of Colorado, Boulder, CO 80309-0389 \\
13. Eureka Scientific, Austin, TX \\
14. Spitzer Science Center, Caltech, Pasadena, CA \\
15. European Southern Observatory, Santiago, Chile \\
}

\begin{abstract}

We analyze the submillimeter emission surrounding the new FU Orionis-type object, HBC 722.  We present the first epoch of observations of the active environs of HBC 722, with imaging and spectroscopy from PACS, SPIRE, and HIFI aboard the Herschel Space Observatory, as well as CO J$=$ 2-1 and 350 $\mu$m imaging (SHARC-II) with the Caltech Submillimeter Observatory.  The primary source of submillimeter continuum emission in the region -- 2MASS  20581767+4353310 -- is located 16$\arcsec$  south-southeast of the optical flaring source while the optical and near-IR emission is dominated by HBC 722.  A bipolar outflow extends over HBC 722; the most likely driver is the submillimeter source.  We detect warm (100 K) and hot (246 K) CO emission in the surrounding region, evidence of outflow-driven heating in the vicinity.  The region around HBC 722 itself shows little evidence of heating driven by the new outbursting source itself.

\end{abstract}

\section{Introduction}

FU Orionis-type objects (hereafter, FUors) are a class of low-mass pre-main
sequence objects named after the prototype FU Orionis, which
produced a 5 magnitude optical outburst in 1936 and has remained in its
brightened state ever since.  About five more sources have been observed to flare prior to 2010;
similar spectral characteristics (broad blueshifted emission lines, IR excess, near-IR CO
overtone absorption) have been used to identify $\sim$ 10 additional
FUor-like objects \citep{hartmann96a, reipurth10}.  \citet{hartmann96a} proposed that FUors are the result of a sudden cataclysmic accretion of material from a reservoir that had built up in the circumstellar disk surrounding a young stellar object.  Models of FUor outbursts indicate that over only a few months the accretion rate rises from the typical rate for a T Tauri star ($\dot{M}$ $\lesssim$
10$^{-7}$ M$_{\odot}$ yr$^{-1}$) up to 
10$^{-4}\,$M$_{\odot}$ yr$^{-1}$, and then decays 
over an e-fold time of $\sim$ 10-100 yr \citep{bell94}.  Over the entire outburst the star could accrete 
$\sim$ 0.01 M$_{\odot}$ of material, roughly the mass of a typical T
Tauri disk \citep{andrews05}.  Mid-infrared observations of FUors with the Spitzer Space Telescope reveal that some resemble embedded protostars, in which their excess emission at wavelengths greater than $\sim$ 30 $\mu$m cannot be accounted for by circumstellar disk models alone, while others lack a strong mid-IR excess above that of a mildly flared disk \citep[e.g.][]{green06,quanz07a,zhu08}.  FUors are connected with both T Tauri stars and more embedded systems, repeating outburst cycles with replenished material from an infalling envelope \citep{weintraub91,sandell01,quanz07a,zhu10}.  Until now we have not had the opportunity to study the far-infrared emission from an FUor close to maximum light, within a few months of outburst.  

HBC 722, also known as LkH$\alpha$ 188-G4 and PTF10qpf, is located in the ``Gulf of Mexico'' in the southern region of the North American/Pelican Nebula \citep[520 pc distance;][comparable to FU Orionis itself at 500 pc]{straizys89, straizys93, laugalys06}.  HBC 722 (20h58m17.0s +43d53m42.9s J2000) was identified independently as an FUor by \citet{semkov10b} and \citet{miller10short}. Prior to outburst it was identified  as an emission-line object of spectral type K7-M0 by \citet{cohen79}, who estimated an interstellar reddening of A$_V$ $=$ 3.4 mag.  The rise in luminosity began slowly, but accelerated during the summer of 2010 \citep{semkov10a} and was monitored in the optical and near-IR \citep{semkov10b,miller10short, munari10, leoni10, kospal11short}, exhibiting two separate epochs of behavior: during the June-early August 2010 period it rose $\sim$ 1 mag (beginning at 16.5 mag in R-band in 2009); then it increased 3 mag in the next 50 days, before reaching peak flux in late September 2010.  It has since declined by 0.5 mag at optical wavelengths; this decline rate would suggest a return to quiescence within 1 yr \citep{kospal11short}, considerably faster than a classical FUor; however from AAVSO photometry\footnote{www.aavso.org, request of C. Aspin} it appears that the rate of decay has since slowed.  No FUor has been as well-characterized prior to outburst as HBC 722.

\section{Observations and data reduction}

HBC 722 was observed with Herschel as a Target of Opportunity (PI: J. Green) during Regular Science Phase \citep{pilbratt10short}.   The 50$\arcsec$$\times$50$\arcsec$  region around HBC 722 was observed with all three spectroscopic instruments on Herschel (PACS, SPIRE, and HIFI), and with the 230 GHz heterodyne receiver and SHARC-II \citep{dowell03short} on the Caltech Submillimeter Observatory (CSO).  All of these observations were obtained between Nov. 30 and Dec. 14, 2010, and the Herschel observations are planned to be repeated six months later.

\subsection{Imaging}

\begin{figure}
\begin{center}
\includegraphics[scale=0.6, angle=0]{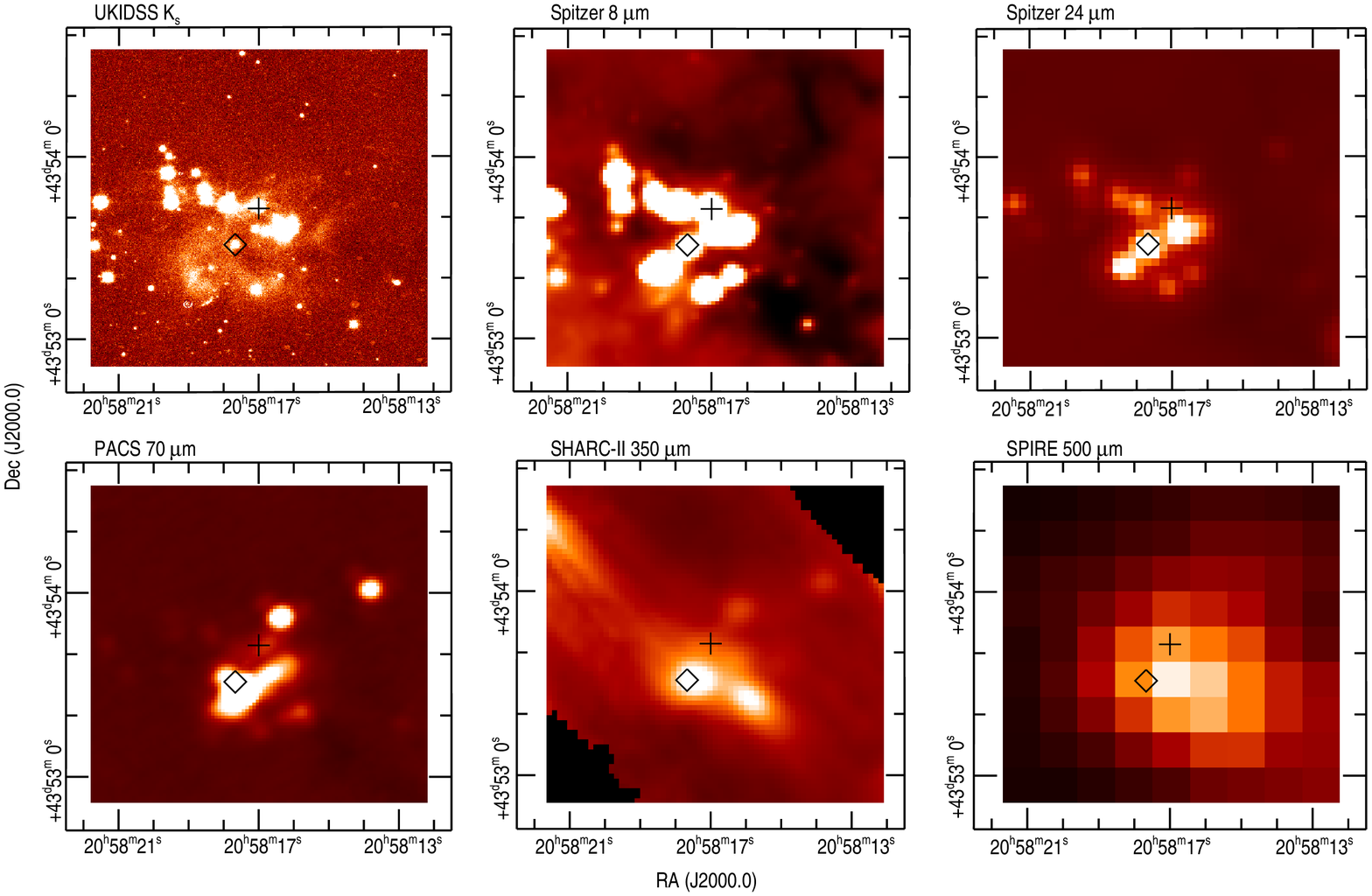}
\caption{The region surrounding HBC 722 (+ symbol) and 2MASS 20581767+4353310 (diamond). Top (pre-outburst), left: K$_{\rm S}$-band (UKIRT); center: 8 $\mu$m \citep[IRAC;][]{guieu09short}; right: 24 $\mu$m \citep[MIPS;][]{rebull11short}.  Bottom (post-outburst), from left to right:  70 $\mu$m (PACS; this work) 350 $\mu$m (SHARC-II; this work), 500 $\mu$m (SPIRE; this work).  The offset between HBC 722 and the 2MASS source is $\sim$ 16$\arcsec$.  The offset in the 500 $\mu$m image is real; the westernmost source in the 350 $\mu$m image has an even redder spectrum than 2MASS 20581767+4353310, and dominates the flux at wavelengths greater than 350 $\mu$m. }
\label{hbc722_spirephot}
\end{center}
\end{figure}

K-band, Spitzer-IRAC 8 $\mu$m, Spitzer-MIPS 24 $\mu$m pre-ouburst, PACS, SPIRE, and SHARC-II post-outburst images of HBC 722 and its surroundings are shown in Figure \ref{hbc722_spirephot}.  The K$_S$-band image was obtained as part of the UKIRT Infrared Deep Sky Survey (UKIDSS), taken with the Wide Field Camera on the 3.8 m diameter UKIRT
in 2006, as part of Data Release 8.  The Spitzer data were obtained as part of Programs 20015 and 462 (P.I.: L. Rebull).  PACS photometry was gathered in single-cycle integration times in the small scan map mode at orientation angles of 70$^{\circ}$ and 110$^{\circ}$.  The beam sizes at 70, 100, and 160 $\mu$m are approximately 5.5$\arcsec$, 6.5$\arcsec$, and 11$\arcsec$, respectively.  SPIRE photometry was gathered in single-cycle integration times in the small map mode.  SPIRE observes simultaneously at 250, 350 and 500 $\mu$m. The on-orbit beam sizes are 18.1$\arcsec$, 25.2$\arcsec$, and 36.6$\arcsec$, respectively.   The Herschel data were reduced using HIPE \citep[Herschel Interactive Processing Environment;][]{ott10} pipeline v6.0.  The SHARC-II bolometer array \citep{wu07} has a 2.9\arcmin$\times$0.7$\arcmin$ field of view and a beamsize of 8.5$\arcsec$  at 350 $\mu$m, and $\tau_{225~GHz}$ was stable at $\sim$ 0.03.

\subsection{Spectroscopy}

HIFI \citep{degraauw10short} was used in single point mode, in band 6b (1578.2--1697.8 GHz) tuned to CO J$=$14-13 (1611.8 GHz), and in band 1b (562.6--628.4 GHz) for CO J$=$5-4 (576.3 GHz).  The data were again reduced using HIPE.  The beam size of HIFI is 43.1$\arcsec$  in band 1b, and $\sim$ 15$\arcsec$  in band 6b; thus the low-J lines contain far more emission from the submillimeter source, while the high-J lines are less confused.

SPIRE-FTS \citep{griffin10short} was utilized as a single pointing with sparse image sampling, high spectral resolution, in 1 hr of integration time.  The spectrum is divided into two orders covering the spectral ranges 194 -- 325 $\mu$m (SSW) and 320 -- 690 $\mu$m (SLW), with a resolution of $\lambda$/$\Delta\lambda$ $\sim$ 300--800, increasing at shorter wavelengths.  Each order is reduced separately within HIPE using the standard pipeline for extended sources, including apodization to remove the Airy diffraction pattern.  None of the lines in the HBC 722 vicinity detected with SPIRE are spectrally or spatially resolved, and thus include components from all sources in the region.  The measured rms of the spectra range from 0.2 Jy to 0.6 Jy, as expected from the Herschel time estimator.  We recover the same total CO J$=$ 5-4 flux measured with HIFI.  Additionally, most lines appear blueshifted by $\sim$ 200 km s$^{-1}$, likely an artifact of the pipeline reduction for SPIRE-FTS; the velocity-resolved HIFI data do not show this effect, nor do the PACS data.

PACS \citep{poglitsch10short} is a 5$\times$5 array of 9.4$\arcsec$ $\times$9.4$\arcsec$  spatial pixels (also referred to as ``spaxels'') covering the spectral range from 51 -- 210 $\mu$m with a resolution $\lambda$/$\Delta\lambda$ $\sim$ 1000--3000, divided into four orders, covering $\sim$ 50--75, 70--105, 100--145, and 140--210 $\mu$m.  We reached expected rms sensitivity of $\sim$ a few $\times$ 0.1 Jy (continuum) or $\sim$ 3-10 $\times$ 10$^{-18}$ W m$^{-2}$.  Each order was first reduced using a modified pipeline based on the standard PACS spectral reduction script within HIPE.  The relative spectral response function within each band was determined both before launch and modified using calibration data taken in the early phases of the mission.  Further improvements in the flatfielding, spectral response, and defringing are underway and will further improve the spectral shape; however line fluxes are not likely to be affected strongly except for a reduction in the rms uncertainty.  The telescope sky background was subtracted using two nod positions 6\arcmin\ from the source.  The absolute flux uncertainty is estimated at 20\%.

\section{Analysis}

\subsection{HBC 722}

From the photometry it is clear that the position corresponding to the outbursting source does not emit strongly in continuum at wavelengths greater than 70 $\mu$m, consistent with the low extinction and disk-like spectrum measured pre-outburst.  At 70 $\mu$m we detect a point-source with F$_{\nu}=$ 0.412 $\pm$ 0.020 Jy, but this value should be considered an upper limit as it includes co-spatial cloud emission.  Like many other FUors, the flaring source has an offset submillimeter counterpart \citep{evans94,sandell01}.  In order to isolate the gas component due to HBC 722, which is only possible in the PACS and CSO data due to spatial resolution, we extracted the central spaxel separately, which encompasses a 10$\arcsec$ square covering HBC 722 itself.  The J$=$14-13 to J$=$21-20 CO emission from the spaxel that includes HBC 722 is fit with high uncertainty by a single temperature (289 K $\pm$ 212 K).  If the CO temperature toward HBC 722 is raised by the outburst, this effect may increase during the second epoch of observations six months later.

\begin{figure}
\begin{center}
\includegraphics[scale=1.0,angle=270]{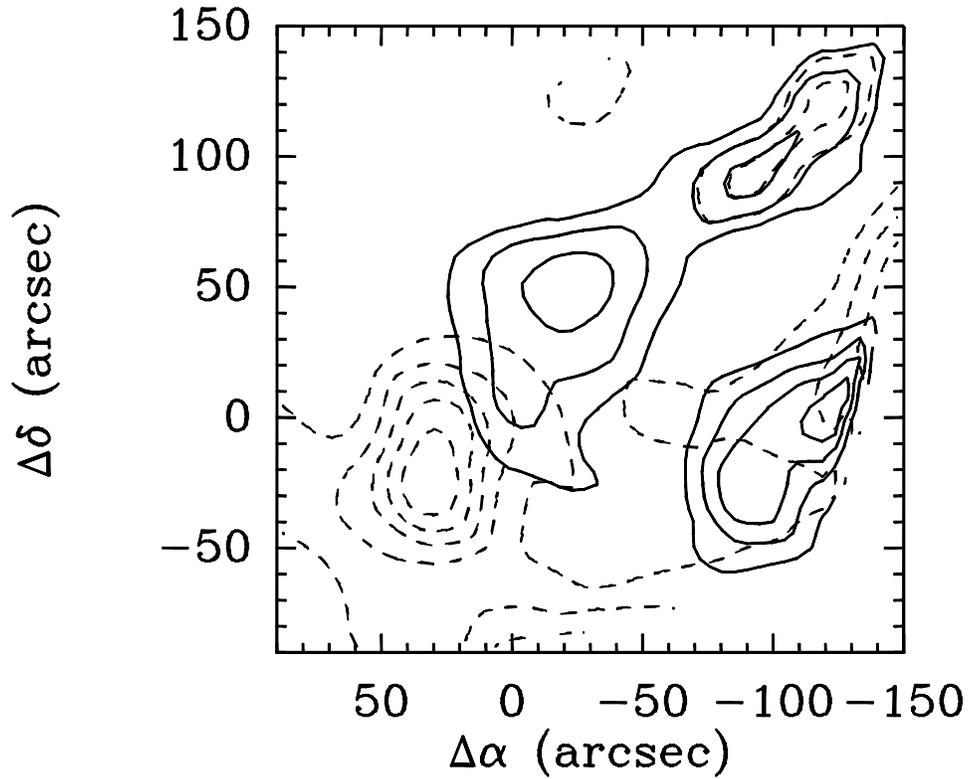}
\caption{CSO map of CO J$=$ 2-1 of the HBC 722 vicinity. The position of the optical/near-IR outburst is (0,0).
The map spacing is 30\arcsec, roughly equal to the beam size. The
velocity ranges are $-5$--0 km s$^{-1}$\ for the blue wing (dashed) and 9--14
km s$^{-1}$\ for the red wing (solid). Contours are 1--5 K km s$^{-1}$ for the
blue and 1--4 K km s$^{-1}$ for the red, spaced by 1 K km s$^{-1}$, about
2.5 $\sigma$. There are clearly several outflows in the region.}
\label{csomap}
\end{center}
\end{figure}

Figure \ref{csomap} shows the distribution of CO J$=$ 2-1, in blue and redshifted contours, showing a clear bipolar structure projected on or behind HBC 722 as well as several additional outflows; the spatial resolution (32$\arcsec$) is insufficient to identify the driving source, but HBC 722 clearly sits in an active star forming region.  The FWHM of the J$=$ 5-4 line observed with HIFI is $\sim$ 0.04 GHz, or 7.45 km s$^{-1}$, centered at 1611.7582 GHz, consistent with the CO J$=$ 2-1 spectra.  The profile of this line is almost exactly matched to the J$=$ 2-1 line -- barring the stronger absorption in the J$=$ 5-4 profile, indicating a similar emitting source; however it is quite strong compared even to most embedded protostars \citep[e.g.][]{yildiz10short}; furthermore the SPIRE-FTS measurement of the J$=$ 5-4 line matches the velocity-integrated HIFI flux.  This further indicates that the SPIRE spectrum is dominated by a stronger submillimeter source, while the PACS central spaxel is characterizing HBC 722 itself.  It is also possible that {\it all} of the CO emission at this location may be due to outflows unrelated to HBC 722.

\subsection{2MASS 20581767+4353310}

The SHARC 350 $\mu$m image resolves the SPIRE photometry at the same wavelength range into multiple sources, clearly delineating the dominant submm source.  The submillimeter peak is strong but offset from the location of the FUor by $\sim$ 16$\arcsec$, instead aligning with a different and much redder source, identified with the 2MASS source 20581767+4353310  \citep[also identified as JCMTSF J205817.0+435327, or 205817.66+435331.0;][]{rebull11short}.  This situation is similar to the RNO 1b/c region that harbors two FUor-like objects and more deeply embedded sources nearby \citep{quanz07b}.  This second source appears as a minor but clearly separate source between 1.25 and 8 $\mu$m, but dominates the flux at wavelengths $\ge$ 24 $\mu$m \citep{guieu09short,rebull11short} as well as the PACS and SPIRE photometric bands, and the SHARC 350 $\mu$m image.

\begin{figure}
\begin{center}
\includegraphics[scale=0.8, angle=0]{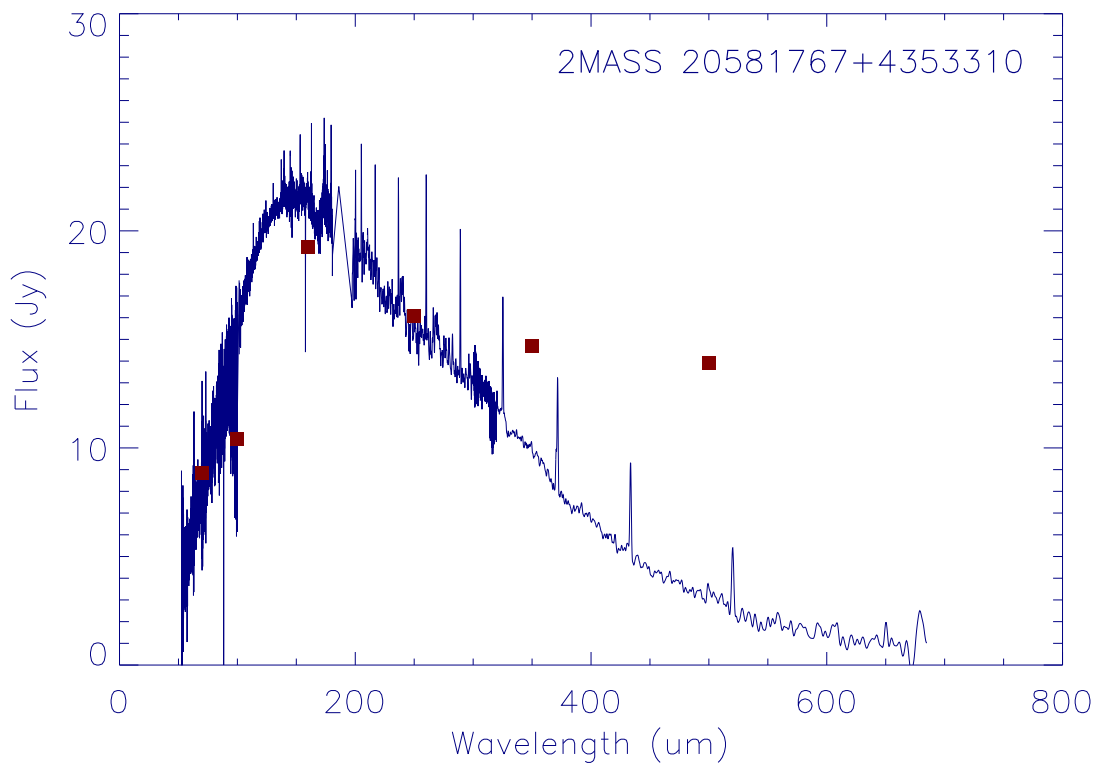}
\caption{PACS and SPIRE-FTS 52 - 690 $\mu$m spectrum and photometry (red squares) of 2MASS 20581767+4353310.  The PACS spectrum is extracted from a 2$\times$2 group of spaxels surrounding the source, roughly the size of the SPIRE beam at shorter wavelengths.  The SPIRE spectrum was scaled downward by a factor of 1.5 to match.  Even with this adjustment it is likely that the SPIRE spectrum includes additional flux from other embedded nearby sources compared to the PACS spectrum.  The drop at 100 $\mu$m is an order edge artifact in PACS band B2B.  The narrow emission features are unresolved lines.  The narrow absorption features ([N III], [O III], and [C II]) are spurious and introduced by emission lines at the nod positions.  The SPIRE photometry at 500 $\mu$m includes both 2MASS 20581767+453310 and the additional western source.}
\label{hbc722_spirespec}
\end{center}
\end{figure}

We extracted 2$\times$2 spaxel areas centered on the best-known location of 2MASS 20581767+4353310.  The submm source is off-center to the south by 2 spaxels, or 16$\arcsec$, another source of absolute flux uncertainty.   The SPIRE-FTS spectrum of the region is rich in gas lines, including CO J$=$5-4 up to J$=$13-12, [C I] at 370.4 $\mu$m (blended with the CO J$=$ 7-6 line), and [N II] at 205 $\mu$m, a likely outflow indicator seen in other FUors \citep{lorenzetti00short}.  In the PACS spectrum we can clearly see that the continuum emission and much of the line emission is focused near the position of the submillimeter source, 2MASS 20581767+4353310, and not HBC 722.  We detected a wealth of high-J CO emission up to J$=$ 21-20, [O I], but no indication of [N II], [N III], or [O III]; in fact, [O III] 88.3 $\mu$m and [N III] 57.3 $\mu$m, and [C II] 157 $\mu$m appear in absorption due to the superposition of an HII region in the off-positions.   The combined PACS-SPIRE spectrum of 2MASS 20581767+4353310 is displayed in Figure \ref{hbc722_spirespec}, with the SPIRE spectrum scaled downward by a factor of 1.5 to match, and the detected lines are listed in Table \ref{hbc722_spirelines}. The source has brightened in the past decade; the UKIRT (observed in 2006) JHK magnitudes are $\sim$ 0.2 mag higher than the 2MASS (observed in 2000) values (the photometric uncertainties are $\sim$ 0.036-0.054 mag for 2MASS, and $\sim$ 0.07-0.09 for UKIRT).  A secondary source visible to the west in the SHARC image that appears at wavelengths beyond 250 $\mu$m may be contributing as well, and although it has no 2MASS counterpart it is visible in the UKIRT image, separated from 2MASS 20581767+4353310 by an extinguished region.

\begin{figure}
\begin{center}
\includegraphics[scale=0.9, angle=0]{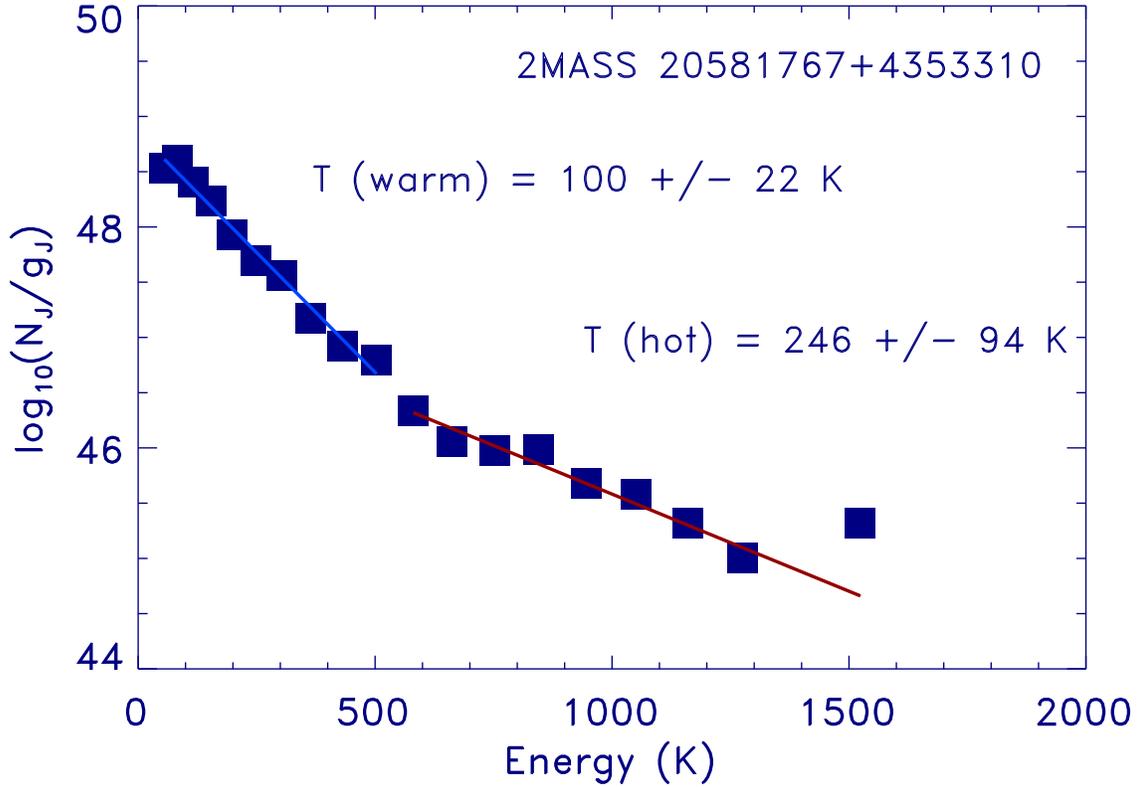}
\caption{CO Rotational diagram for 2MASS 20581767+4353310, covering from J$=$ 5-4 up to J$=$ 23-22, fit with two temperature components.  The vertical axis is the log of the number of molecules (assuming optically thin emission) in state J divided by the degeneracy of the state.  The SPIRE points are scaled downward by 1.5 to match the PACS continuum flux.  The J$=$ 23-22 (1520 K) line is partly blended with H$_2$O, and the J$=$ 22-21 line is not detected; both are ignored in the fit.}
\label{corotdiagram}
\end{center}
\end{figure}

The region around 2MASS 20581767+4353310 shows a rich gas spectrum indicative of heated ambient material.  A single temperature fit to the full CO rotational diagram does not characterize the data very well; a break in the data points occurs around E$_u$/k$=$ 600 K.   The lower energy points yield a temperature of 100 $\pm$ 22 K, while the higher energy points indicate 246 $\pm$ 94 K (Figure \ref{corotdiagram}).   Optical depth effects would lead to an underestimate of the population in lower J levels, causing a deviation below the fit at low energy. We see no obvious sign of this effect.
Although the rotational diagram is uncertain due to source confusion at the spatial resolution of Herschel and the combined use of SPIRE and PACS, some of these problems are mitigated:  the choice of breakpoint is close to the boundary between the instruments but the results do not change significantly if the break is adjusted slightly in either direction; the extracted region in PACS and SPIRE is similar in spatial extent; lastly, we are more likely to underestimate the flux from the extended
outflow (Figure \ref{csomap}) at the shortest wavelengths where the beam is smallest, causing a drop in the high energy state populations.  We do not observe this effect; if it is significant, the evidence for the
higher temperature component becomes stronger.

Models of embedded sources such as HH46 \citep[e.g.][Visser et al. 2011, in prep.]{vankempen10short, yildiz10short} generate $\sim$ 100 K CO emission through passive heating while 250 K gas might originate from UV heating; an $\sim$ 800-1000 K component that would be generated by shocks is {\it not} detected here.  We detect low-lying states of H$_2$O emission, which can be indicative of outflows and photon-heating, although no OH is detected.  We derive from the CO rotational diagram N$_{tot}=$ 2.58 $\pm$ 0.48 $\times$ 10$^{50}$, which for an emitting area of 40$\arcsec$  at the proposed distance of 520 pc yields a column density of 9.40 $\times$ 10$^{14}$ cm$^{-2}$, broadly consistent with other embedded sources observed with Herschel \citep[e.g.][]{vankempen10short}.  Measured from only the Herschel spectrum and photometry centered on this source L$_{Bol}=$ 5.4 $\pm$ 1.8 $\lsun$ and T$_{Bol}=$  33.4 $\pm$ 11.0 K; clearly it and the other western source dominate the submillimeter emission while contributing little to the optical/near-IR portion of the spectrum relative to HBC 722 itself.

\subsection{Discussion}

Given the recent nature of the outburst, it is possible that any shocks from HBC 722 have not yet had time to propagate into a large enough area to be observed.  A hot disk of radius 0.5 AU would have a very small beam filling factor of 3.3 $\times$ 10$^{-9}$ in band 6b.  If instead the emission is driven by a newly launched jet, moving at 200 km s$^{-1}$, the ejecta could have traveled $\sim$ 10 AU in the three months between outburst and our Herschel observations; even if the outflow is spherically symmetric, the filling factor is capped at 1.3 $\times$ 10$^{-6}$.  Most likely the observed CO originates in the older outflow traced in the submillimeter line maps, which is extended at least 50$\arcsec$, and therefore is not associated with the present outburst.  However, UV photons emitted during the outburst will have traveled $\sim$ 15$\arcsec$ in this time; therefore if HBC 722 is within this distance of the outflow cavity walls, we may expect that the CO flux will increase in our second observation as the UV photons will have traveled $\sim$ four times the distance, encompassing an area 16 times greater and exciting high-J CO lines.

The 100 K CO traces PDR emission (presumably from jets) from the cold gas around embedded sources, as well as outflows in the region.  We might attribute [N II] (205 $\mu$m) emission to a dissociative shock from a newly launched jet, but it may also be tracing the older outflow; however the absence of [N II] (122 $\mu$m) as well as more highly ionized lines ([N III] and [O III]) suggest that any jet emission is not contributing significantly.  If during the second epoch the jet has excited the expanded area, highly ionized emission may become detectable.  At this time we see little evidence for shocks but due to the recent nature of the outburst it is possible that the filling factor of any newly launched jet material has not yet risen sufficiently to detect their signature.  Two complementary Herschel open time programs to observe older FUors (PIs: J. Green; M. Audard) will provide a comparison with sources that have decayed from their respective peaks for decades.

\section{Conclusions}

We present the full suite of Herschel observations of HBC 722 and the surrounding region.  HBC 722 does not show evidence for a circumstellar envelope or shocked gas, and appears to have erupted from a disklike state,  similar to FU Orionis itself.  The far-IR/submm emission in this region is dominated by nearby sources such as 2MASS 20581767+4353310, rather than the counterpart to the visible/near-IR FUor.   In later epochs the effect of the HBC 722 outburst may become clear through increased heating of CO, or emission from shocked gas.  We will re-observe the region with Herschel in May 2011.

\acknowledgements

The authors wish to acknowledge the Herschel Director, G. L. Pilbratt, for the timely approval and execution of this Target of Opportunity program; Amanda Heiderman for providing CSO data; Paul Harvey, Michelle Rascati, and the HSC and NHSC Helpdesk for data processing and analysis assistance, and Colette Salyk and Geoff Blake for helpful discussions.  The research of \'A.K. is supported by the Nederlands Organization for
Scientific Research.  The research of J.-E. L. is supported by Basic Science Research Program through the National Research
Foundation of Korea (NRF) funded by the Ministry of Education, Science and Technology (No.
2010-0008704).  This work is based in part on data obtained as part of the UKIRT
Infrared Deep Sky Survey.



\begin{center}
\begin{deluxetable}{l r r r r r r r r r}
\tabletypesize{\scriptsize}
\tablecaption{Spectral Lines Detected in 2MASS 20581767+4353310  \label{hbc722_spirelines}}
\tablewidth{0pt}
\tablehead{
\colhead{Line} & \colhead{Frequency} & \colhead{Wave (lab)} & \colhead{Wave (obs)} & Width & \colhead{Flux} & \colhead{RMS} & \colhead{E$_{up}$} & \colhead{A-Coef.} & \colhead{State Deg.}}
 \startdata
 & &  & & &10$^{-18}$ &10$^{-18}$ & & & \\
 Units: & GHz &$\mu$m  & $\mu$m & $\mu$m & W m$^{-2}$ & W m$^{-2}$ & K & s$^{-1}$ & \\
 & & & & & & & \\
CO J $=$ 4-3 & 461.041 & 650.8 & 650.26 & 2.18 & 17.61 & 2.95 & 55.32 & 6.126e-06 & 9 \\
CO J $=$ 5-4 &  576.27 & 520.6 & 520.18 & 1.83 & 63.35 & 4.40 & 82.97 & 1.221e-05  &  11 \\
CO J $=$ 6-5 & 691.47 & 433.9 & 433.53 & 1.33 & 100.65 & 3.52 & 116.16 & 2.137e-05 & 13 \\
CO J $=$ 7-6 & 806.65 & 371.9 & 371.62 & 1.19 & 146.13 & 3.66 & 154.87 & 3.422e-05 & 15 \\
$[$C I] $^3$P$_1$-$^3$P$_2$ & 809.94 & 370.4 & 370.5 & 0.97 & 46.54 & 3.05 & -- & -- & -- \\
CO J $=$ 8-7 & 921.80 & 325.5 & 325.23 & 0.82 & 141.43 & 4.14 & 199.11 & 5.134e-05 & 17 \\
CO J $=$ 9-8 & 1036.91 & 289.3 & 289.15 & 0.55 & 145.75 & 4.71 & 248.88 & 7.33e-05 & 19 \\
CO J $=$ 10-9 & 1151.99 & 260.4 & 260.21 & 0.49 & 180.14 & 9.77 & 304.16 & 1.006e-04 & 21 \\
CO J $=$ 11-10 & 1267.01 & 236.8 & 236.63 & 0.36 & 120.13 & 6.45 & 364.97 & 1.339e-04 & 23 \\
CO J $=$ 12-11 & 1382.00 & 217.1 & 216.94 & 0.29 &103.34 & 7.89 & 431.29 & 1.735e-04 & 25 \\
$[$N II] $^3$P$_1$-$^3$P$_0$ & 1462.14 & 205.18 & 205.17 & 0.33 & 123.51 & 9.02 & 14.53 eV & -- & -- \\
CO J $=$ 13-12 & 1496.92 & 200.4 & 200.17 & 0.43 & 114.12 & 12.99 & 503.13 & 2.200e-04 & 27 \\
CO J $=$ 14-13 & 1611.79 & 186.00 & 186.01 & 0.13 & 57.45 & 6.92 & 580.49 & 2.739e-04 & 29 \\
o-H$_2$O 2$_{12}$-1$_{01}$ & 1671.06 & 179.53 & 179.53 &  0.15  &  58.84 &  8.51 & 114.4 & 5.593d-02 & 15 \\
CO J $=$ 15-14 & 1726.60 & 173.63 & 173.65 & 0.11 & 42.11 & 7.01 & 663.35 & 3.354e-04 & 31 \\
CO J $=$ 16-15 & 1841.35 & 162.81 & 162.82 & 0.08 & 47.86 & 5.09 & 751.72 & 4.050e-04 & 33 \\
CO J $=$ 17-16 & 1956.02 & 153.27 & 153.27 & 0.17 & 65.99 & 9.37 & 845.59 & 4.829e-04 & 35 \\
$[$O I] $^3$P$_0$-$^3$P$_1$ & 2061.42 & 145.53 & 145.53 & 0.15 & 31.62 & 12.11 & -- & -- & -- \\
CO J $=$ 18-17 & 2070.62 & 144.78 & 144.80 & 0.11 & 42.48 & 6.94 & 944.97 & 5.695e-04 & 37 \\
p-H$_2$O 3$_{13}$-2$_{02}$ & 2165.63 & 138.53 & 138.60 & 0.17 & 22.58 &  7.41 &  204.7 & 1.251e-01 & 7 \\
CO J $=$ 19-18 & 2185.13 & 137.20 & 137.24 & 0.11 & 44.09 & 4.88 & 1049.84 & 6.650e-04 & 39 \\
CO J $=$ 20-19 & 2299.57 & 130.37 & 130.41 & 0.11 & 30.70 & 5.03 & 1160.20 & 7.695e-04 & 41 \\
p-H$_2$O 4$_{04}$-3$_{13}$ & 2393.23 & 125.35 & 125.41 & 0.25 & 32.99  & 9.23 &  319.5 & 1.727d-01 &  9 \\
CO J $=$ 21-20 & 2413.92 & 124.19 & 124.19 & 0.08 & 18.84 & 4.03 & 1276.05 & 8.833e-04 & 43 \\
o-H$_2$O 4$_{14}$-3$_{03}$$^1$ & 2642.30 & 113.54 & 113.50 & 0.13 & 60.80  &  24.76 &  323.5 & 2.463e-01 & 27 \\
CO J $=$ 23-22$^1$ & 2642.33 & 113.46 & 113.50 & 0.13 & 60.80 & 6.55 & 1524.19 & 1.139e-03 & 47 \\
o-H$_2$O 2$_{21}$-1$_{10}$ & 2758.37 & 108.07 & 108.11 & 0.13 & 38.89  &  6.55 &  323.5 & 2.463e-01 & 27 \\
$[$O I] $^3$P$_1$-$^3$P$_2$ & 4748.33 & 63.18 & 63.19 & 0.06 & 294.30 & 29.4 & -- & -- & -- \\
\\
3$\sigma$ Upper Limits\\
o-H$_2$O 1$_{10}$ -1$_{01}$ & 556.93607 & 538.66 & -- & (1.84) & (12.64) & -- & 61.0  & 3.458e-03 & 9 \\
p-H$_2$O 2$_{11}$ -2$_{02}$ & 752.03323 & 398.92 & -- & (1.33) & (11.72) & -- & 136.9
 &7.062e-03 & -- \\
p-H$_2$O 1$_{10}$ -1$_{01}$ & 1113.34306 & 269.46 & -- & (0.48) & (19.68) & -- &  53.4 & 1.842e-02 & 3 \\
\enddata
\tablecomments{Spectral lines detected with PACS and SPIRE. Lines of wavelength greater than 200 $\mu$m were detected in SPIRE, and less than 200 $\mu$m in PACS.  Source coordinates: 20h58m17.5s + 43d53m28s (closest 2MASS source: 20581767+4353310+4353310).\\
$^1$Blended lines.}
\end{deluxetable}
\end{center}

\end{document}